\documentclass[aps,prb,twocolumn,showpacs,preprintnumbers,amsmath,amssymb,floatfix]{revtex4-1}
\usepackage{graphicx}
\usepackage{dcolumn}
\usepackage{bm}
\usepackage{color}
\usepackage{epstopdf}%
\usepackage{amsmath}%
\usepackage{amsfonts}%
\usepackage{amssymb}

\usepackage{graphicx}
\usepackage{color}

\begin{document}

\title{Robust conductance zeroes in graphene quantum dots and other bipartite systems}
\author{M. Ni\c t\u a}
\author{M. \c Tolea}
\affiliation{National Institute of Materials Physics, Atomistilor 405A,
Magurele 077125, Romania}
\author{D. C. Marinescu}
\affiliation{Department of Physics, Clemson University, Clemson, SC 29634}

\begin{abstract}
%
 Within the Landauer transport formalism we demonstrate that conductance zeroes are possible in bipartite systems at half-filling when leads are contacted to different sublattice sites.  In particular, we investigate the application of this theory to graphene quantum dots with leads in the armchair configuration. The obtained conductance cancellation is robust in the presence of any single-site impurity.
\end{abstract}
\date{\today}
\maketitle

\section{Introduction}

The cancellation of the electronic conductance on account of destructive quantum interference (DQI), independent on the coupling strength to the leads, is a quantum mechanical effect without correspondence in
classical circuits. Finding systems where such property occurs is of both fundamental and practical interest,
 as in designing of on/off switches, for example. The existence of DQI phenomena has been investigated previously in various quantum dots or molecular systems
\cite{tsuji2018,lambert2018,timoti2016,aradhya2013,rotter2005,markussen2010,solomon2015,sam2017,nozaki2017}.
More recently, this topic received renewed attention in connection with the transmission phase lapse of $\pi$ at the conductance zeroes between the resonances of a quantum dot, arguably one of the longest standing puzzle in mesoscopic physics, whose elucidation spanned thirty years \cite{Schuster,Edlbauer}.

In this paper we demonstrate the presence of a robust zero transmission in graphene quantum dots (QD)
at half-filling (i.e. {\it zero} Fermi energy) starting from an analysis of quantum transport in bipartite lattices.
 Such systems, known to provide an appropriate description for graphene, are composed of two sublattices $A$ and $B$ with hopping only between $A$ and $B$ sites and no hopping in the same sub-lattice (see Fig.1). In the Landauer formalism, where the conductance between two points ${\bf G}_{ij}$ is proportional to the transmittance ${\bf T}_{ij}$, it was previously found that zeroes are obtained in graphene QDs when
  both leads are connected to the same sub-lattice, ${\bf T}_{AA}$ or ${\bf T}_{BB}$\cite{tada2002,nita-rrl}.
Moreover, it was shown that this type of zeroes occurs with a
$\pi$ phase lapse of the transmission amplitude, a property characteristic to Fano zeroes.
Here we focus on the origin of the transmission zeroes
and their characteristic properties
in a setup that involves DQI when the transport
leads are connected to both sublattices,
${\bf T}_{AB}$.

{To this end
we first prove the conductance cancellations in a {\it multi-terminal} bipartite conductor
whose transport leads are
contacted to $A$ points.
In some specific circumstances, ${\bf T}_{AA} = 0$ between {\it any} pair of $A$ leads, a result that is left invariant by the presence of a perturbation at {\it any} $A$ sites.} Later, this property is used as a building block in constructing new connected systems, also bipartite, in which the existence of ${\bf T}_{AB}$ zeros is studied. Our theory is then applied to a graphene quantum dot at half-filling, when the two leads are connected to arm-chair edges. The robustness of such conductance zeroes is studied in the presence of lattice defects.

\section{The Landauer Formalism}

\begin{figure}[h]
\centering
\includegraphics[scale=0.72]{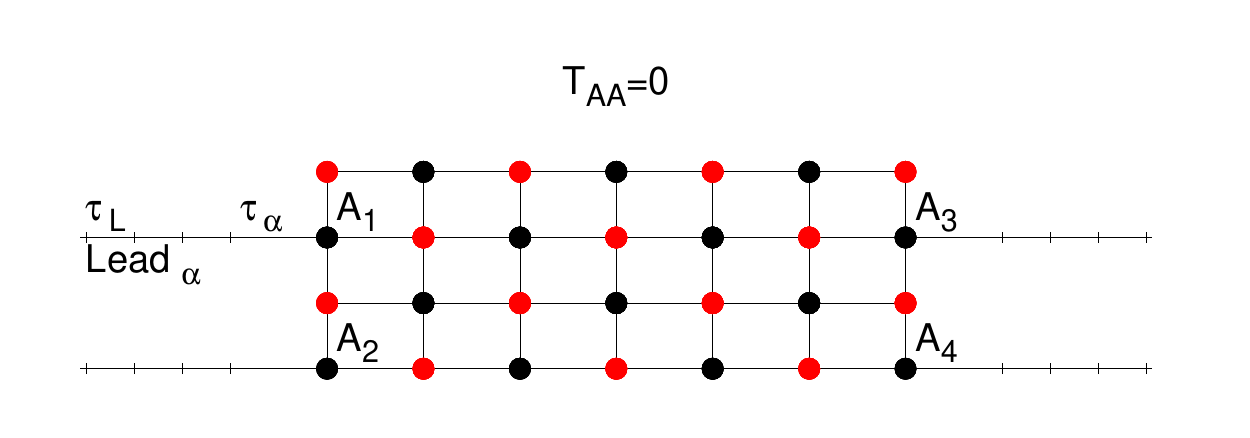}
\caption{
A zero transport bipartite conductor with all terminals
connected at $A$ sites.
The system exhibits zero transmission at $E=0$: ${\bf t}_{\alpha, \bar \alpha}(0)=0$
for any pair of leads $\alpha \ne \bar \alpha$. This is realized when the bipartite Hamiltonian is nonsingular such that it has no zero energy eigenstate. The $T_{AA}$ zeros are invariant under any $A$ site perturbation
{but they can be modified by $B$ site impurities, as discussed in text}.
}
\label{placheta2}
\end{figure}

\begin{figure}
\centering
\includegraphics[scale=0.7]{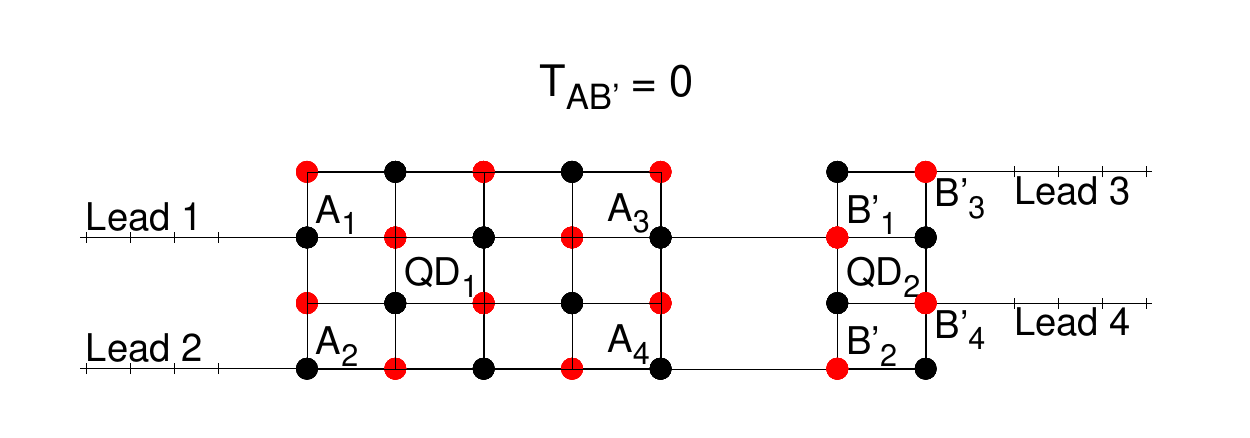}
\caption{
A quantum conductor with $T_{AB}$ zeroes.
We have $T_{13}(0), T_{14}(0), T_{23}(0)$ and $T_{24}(0)=0$. 
The system is composed from two series quantum dots, $QD_1$
and $QD_2$, each of them having zero conductances between any pair of leads as explained in Fig.\,\ref{placheta2}.
The $T_{AB}$ zero is invariant under any perturbation applied to $A$ or $B'$ sites.
Particularly, it can be modified by a selected pair of $B$ and $A'$ sites perturbations.
}
\label{cis}
\end{figure}

The general Hamiltonian of a bipartite lattice considers all the hopping terms between sublattice A and B points,
\begin{equation}\label{hbip}
H =  \sum_{i_A,j_B} t_{i_A, j_B} |i_A\rangle\langle j_B|\;,
\end{equation}
as shown in Figs.\ref{placheta2} and \ref{cis}.

This is a known appropriate representation of nanosized graphene sheets (called also graphene quantum dots)\cite{tada2002,dhakal2019},
artificial molecules composed of connected quantum dots \cite{tamura2002,tolea2016,fernandes2018}
or {alternant} chemical molecules described by the  H\"uckel Hamiltonian \cite{tsuji2018,huckel1933,chen2018}.

In the following considerations we are interested in the general multi-terminal case of a quantum dot (QD) connected to a number of $N_l$ one-channel
transport leads, indexed by $\alpha$ or $\beta=1 \cdots N_l$.
%
%
{\color{black} The leads are described by 1D tight-binding or discrete chain \cite{chen2018, tolea2010, li2019}
and the contact points between them and QD are individual sites denoted with $i_\alpha$ and  $i_\beta$.}

Within the Landauer formalism, the transmission amplitude between leads $\alpha$ and $\bar \alpha$ at
energy E \cite{tolea2010,levi2000},
\begin{equation}\label{tr1}
 {\bf t}_{\alpha, \bar \alpha}  (E)= -\delta_{\alpha, \bar \alpha} 
                          +2 i \frac{\tau_\alpha\tau_{\bar \alpha}}{\tau_l} \sin k ~
                          G^{eff}_{i_\alpha, i_{\bar \alpha}}(E),
\end{equation}
determines the conductance between the same leads,
\begin{equation}\label{tra}
{\bf G}_{\alpha, \bar \alpha}(E)=\frac{e^2}{h}{\bf T}_{\alpha, \bar \alpha}(E)
=\frac{e^2}{h}|{\bf t}_{\alpha, \bar \alpha}(E)|^2,
\end{equation}
with ${\bf T}_{\alpha, \bar \alpha}$ the transmittance.
 {The argument of the transmission amplitude is denoted by $\arg t(E)= \phi(E)$.}
Note that in Eq.~(\ref{tr1}) the effective Green's function
\begin{equation}\label{geff}
G^{eff}(E)=\frac{1}{E-H^{eff}}\;,
\end{equation}
depends on the energy $E=2\tau_l \cos k$, with $k$ the wave number. $\tau_l$ is the lead hopping energy and $\tau_\alpha$
the constriction parameter or the hopping energy between QD and lead $\alpha$.
For simplicity, we assume throughout the paper that
$\tau_\alpha=\tau_{\bar \alpha}=\tau_c$.

The effective Hamiltonian that determines Eq.\,\ref{geff} incorporates in addition to the bipartite Hamiltonian, Eq.~\ref{hbip},
the potential at the contacts $V$ such that $H^{eff} = H + V$ {\color{black} lost its hermiticity, with complex terms given by}
\begin{equation}\label{vcontact}
V=\frac{\tau_\alpha^2}{\tau_l}e^{-ik}\sum_{\alpha=1}^{N_l}|i_\alpha\rangle \langle i_\alpha|\;. 
\end{equation}
{\color{black} The non-hermitian $H_{eff}$ has proven to be a useful tool in describing the
transport properties of open mesoscopic systems \cite{ernzerhof2007, ostahie2016}.}

\section{The $\bf T_{AA}$ zeroes}
We apply the formalism described above to the case of a multi-lead quantum conductor, as depicted in Fig.\,\ref{placheta2}.
All the external leads are connected to the same sublattice of the bipartite system, $A$.
 External perturbations may be present at $A$ sites, $\epsilon_i \ne 0$ with $i\in A$.

In this case, we show that the transmission matrix $t_{\alpha, \bar \alpha}$ with $\alpha \ne \bar \alpha$ satisfies,
\begin{eqnarray}\label{fanot}
&& {\bf t}_{\alpha, \bar \alpha}(0)=0~\mbox{with}~i_\alpha, i_{\bar \alpha} \in A,~
\end{eqnarray}
regardless of how many other leads are connected to the same sublattice points $A$.

This result is derived by using the Dyson expansion for the  effective Green's function $G^{eff}(E)$ in Eq.\,\ref{geff}. For the matrix blocks that contain matrix elements between A sites, $G^{eff}_{AA}(E)$ and $G_{AA}(E)$, we write,
\begin{eqnarray}\label{geff2}
G^{eff}_{AA}(E)=G_{AA}(E)+G_{AA}(E) V_A G^{eff}_{AA}(E),~
\end{eqnarray}
where the potential matrix $V_A$ contains only the A sites terms
from Eq.\,\ref{vcontact}
and the A sites impurities, as we have considered.
We note that on account of the chiral symmetry of the Hamiltonian, the matrix elements of the {\it bare} Green's function $G(E)=1/(E-H)$ between points of the same-sublattice
at {\it zero} energy cancel as previously discussed in Refs.~\cite {nita-rrl,tolea2016,deng2014}. Therefore,
\begin{eqnarray}\label{GAA}
G_{i,i'}(0)=0~\mbox{for}~i,i'\in A~\mbox{or}~i,i'\in B\;.~
\end{eqnarray}
With $G_{AA}(0)=0$ in Eq.\,\ref{geff2} and from Eq.\ref{tr1} one obtains the cancellation from Eq\,\ref{fanot}.

 We note that the validity of this result is conditioned by the absence of the eigenvalue $E=0$ from the bipartite lattice spectrum \cite{tsuji2018,nita-rrl} which assures that the perfect conductance cancellation at $E = 0$ occurs {\it between} resonances. Such a ``perfect" {\it zero} is independent of the coupling strength with the leads, since it is decided by the zeroes of the {\it bare} Green's function. In this respect it is different from the usual low conductance between resonances, which is never a perfect {\it zero} and is, in general, coupling-dependent.

{The invariance of $T_{AA}$ zeroes in Fig.\ref{placheta2} to any $A$ site perturbations may be used to explain
destructive interference in the "off" states for naphtalene or perylene when the contact points of  Buttiker probes and source and drain electrodes
 are belonging to the same sublattice \cite{chen2018}.}
{\color{black} Generally, the invariance of the $T_{AA}$ zeroes let the possibility to lift them
only by perturbations acting at least one $B$ site impurity.}

{\color{black} The multiterminal conductor with $T_{AA}=0$ in Fig.\ref{placheta2} can be used to
explain the occurrence of conductance zeros in bigger systems that incorporate it as a building block.
This will prove important in the next section.}

%


\section{The $\bf { T_{AB} }$ zeroes}

 To prove the existence of the transmission zeroes that appear
when the leads are connected to different sublattices, one at an $A$ site and the other at a $B$ site,
we consider a quantum conductor composed of a sequence of two serially connected quantum dots $QD_1$ and $QD_2$, described in Fig.\,\ref{cis}.
Each quantum dot is a bipartite lattice with no zero energy eigenvalue and with all leads connected to the same sublattice points as in Fig.\,\ref{placheta2}.

{\color{black} The first $QD_1$ is described by the bipartite Hamiltonian $H_1(A,B)$
with $A$ and $B$ its two type of points. In the same way, $H_2(A',B')$ describes the $QD_2$.
The coupling potential between the two dots is
such that it links only $A$ points of the first dot with $B'$ points of the second, as depicted Fig.\,\ref{cis}, so we have the Hamiltonian}
$V_{12}(A,B')=|A_3\rangle \langle B'_1| + |A_4\rangle \langle B'_2| + h.c.$.

The resulting Hamiltonian of the composed system $H_1+H_2+V_{12}$ is bipartite too, with $A+A'$ and $B+B'$ designating the two sublattices.

In the composed bipartite system the tunneling amplitude is zero between points in the $A$ and $B'$ sublattices,
\begin{eqnarray}\label{nofanot}
&& {\bf t}_{\alpha, \beta}(0)=0 ~\mbox{with}~i_\alpha\in A~\mbox{and}~i_\beta\in B'.~
\end{eqnarray}
This is the main result of this section and will be proven below.

The effective Hamiltonian $H^{eff}$ that determines the transmission amplitude in the composed system, in agreement with
 Eq.\,\ref{vcontact},
is written as,
\begin{eqnarray}\label{heff12}
 H^{eff}=&&H_1(A,B)+H_2(A',B')\nonumber\\
&&+ V_{12}(A,B')+  V_1(A) + V_2(B'),
\end{eqnarray}
where $H_1$, $H_2$ describe the two independent QDs, while $V_{12}$ describes the coupling between them.
$V_1(A)$ and $V_2(B')$ are the non-hermitian terms from (\ref{vcontact})
associated with the coupling to the leads.

The matrix elements $G^{eff}_{i_\alpha, i_\beta}(0)$ of $G^{eff}_{AB'}(E)$ for
two lattice points $i_\alpha\in A$ and $i_\beta\in B'$ are calculated from
the Dyson equation written for the total interaction potential in Eq.\,(\ref{heff12}),
\begin{eqnarray}
G^{eff}(E)& = & G(E)+G(E)\left(V_{12}(A,B')\right.\nonumber\\
&+&\left.  V_1(A) + V_2(B')\right)G^{eff}(E)\;.\label{geff12-zero}
\end{eqnarray}
Since the initial
system $H$ in (\ref{heff12}) is decoupled, its
Green's function matrices $G_{AB'}$ and $G_{AA'}$ are equal to zero in the expansion of the Dyson equation, leading to
\begin{eqnarray}\label{geff12}
G^{eff}_{AB'}(E)=G_{AA}V^1_{AA}G^{eff}_{AB'}+G_{AA}V^{12}_{AB'}G^{eff}_{B'B'}.
\end{eqnarray}
$V^1_{AA}$ and $V^{12}_{AB'}$ are the matrices of the operators $V_1(A)$ and $V_{12}(A,B')$ in (\ref{heff12}).
Since $QD_1$ as a bipartite system does not have an $E=0$ eigenstate and $G_{AA}(0)=0$ in Eq.\,\ref{GAA}, $G^{eff}_{AB'}(0)=0$. Then, with input from (\ref{tr1})
the cancellation (\ref{nofanot}) follows.

A slightly less general result is obtained by considering a single incoming
and a single outgoing lead. One lead is on an $A$ site coupling to the point $i_\alpha$
and the other lead is at a $B'$ site coupling to the point $i_\beta \in B'$.
For this two-terminals conductor one can prove that
the transmission zero ${\bf t}_{\alpha, \beta}(0)$  has no $\pi$ phase lapse.
In the formula (\ref{geff12}) of $G^{eff}_{AB'}(E)$ we introduce
the Dyson expansion for $G^{eff}_{B'B'}(E)$ and retain only the lowest order term in the limit of $E\to 0_\pm$
when the bare functions $G_{AA}\to 0$ and $G_{B'B'}\to 0$.
We obtain,
\begin{eqnarray}\label{geff3}
G^{eff}_{AB'}(0_\pm) \simeq G_{AA}(0_\pm)V^{12}_{AB'}G_{B'B'}(0_\pm).
\end{eqnarray}
The transmission ${\bf t}_{\alpha, \beta}(0_\pm)$ in Eq.~(\ref{tr1})
becomes a summation of products $G_{i_A, j_A} G_{i_{B'}, j_{B'}}$ with
 $i_A$, $j_A$ $\in A$ and $i_{B'}$, $j_{B'}$ $\in B'$.
Since every product term $G_{i_A, j_A}$ or $G_{i_{B'}, j_{B'}}$ describes a $\pi$ phase lapse process \cite{nita-rrl},
an overall $2\pi$ phase is obtained and consequently no observable phase variation occurs.
From these considerations one obtains:
\begin{eqnarray}\label{delta12}
\Delta \arg {\bf t}_{\alpha \beta}(0)  =  0.
\end{eqnarray}

The stability of the ${\bf T}_{AB}$ zero obtained in (\ref{nofanot}) is now investigated
in the presence of a disorder potential  represented by  impurity energies located at various sites of the lattice.
The effective total Hamiltonian becomes,
\begin{equation}
H'^{eff}=H^{eff}+\sum_i \epsilon_i |i \rangle \langle i |\;.
\end{equation}

From the Dyson expansion for $G'^{eff}_{AB'}$, 
straightforward calculations
lead to
\begin{equation}
G'^{eff}_{AB'}(0)=G^{eff}_{AB}(0) \epsilon_B G'^{eff}_{BA'}(0) \epsilon_{A'} G^{eff}_{A'B'}(0)\;,
\end{equation}
where $\epsilon_B$ and $\epsilon_{A'}$ are the matrices of $B$ and $A'$ located impurities.
 Eq.\,\ref{tr1}
generates the lowest order terms of the tunneling amplitude between contact points $i_\alpha = A_1, A_2$ and $i_\beta =B'_3, B'_4$,
\begin{eqnarray}\label{tpeff}
{\bf t}_{\alpha, \beta}(0)= \epsilon_B  C_{BA'} \epsilon_{A'}+{\cal O}(\epsilon^3).
\end{eqnarray}
$C_{BA'}$ is a matrix containing Green's function products derived by the perturbative method.
{
For instance for output lead connected at $i_\alpha = A_1$ and the input one with $i_\beta =B'_3$ it is written
$C_{BA'}=G_{A_1B}G_{BA}V^{12}_{AB'}G_{B'A'}G_{A'B'_3}$, with all Green functions at E=0 calculated for $H_1+H_2$ from Eq.\,\ref{heff12}.
}

This result shows a significant difference between the ${\bf T}_{AB}$ and ${\bf T}_{AA}$ zeroes.
The cancellation ${\bf T}_{AB'}=0$ in Fig.\,\ref{cis} is invariant in the presence of any single-site impurity and
could be modified only by at least a selected  pair of $A', B$ located impurities.
In contrast, the existence of a same sub-lattice zero, ${\bf T}_{AA}=0$ in Fig.\,\ref{placheta2}, is invariant in the presence of any $A$ site located impurities, but can be modified by one $B$ located impurity.

{\color{black} In this paper we have focused on the non interacting Hamiltonian systems to predict the general features
of the DQI processes \cite{tsuji2018, markussen2010, levi2000}.
At discussed in other works the presence of interaction terms (on site or long range) may lead to the energy shift,
small diminishing or to the energy splitting of the DQI dips \cite{markussen2010, tsuji2014, tsuji2019, valli2019}.
%
%
One may
expect that the obtained DQI processes to be a generic feature even
in the presence of interaction as long as the adiabatic turning on
of the interaction terms do not induce new energy levels (or density
peak) between QI adjacent resonance energies. This is discussed on
the base of the Friedel sum rule in Ref.\,\onlinecite{lee1999}.
We mention also that the electron-hole
symmetry (specific to bipartite lattices) survives interaction
models such as Hubbard or extended Hubbard (such as PPP model \cite{Pariser,Pople}).
Remarkably, electron-hole symmetry was also proven experimentally
in a carbon nanotube \cite{Herrero}.
}


\section{The arm-chair zeroes in graphene}
\begin{figure}
\centering
\includegraphics[scale=0.6]{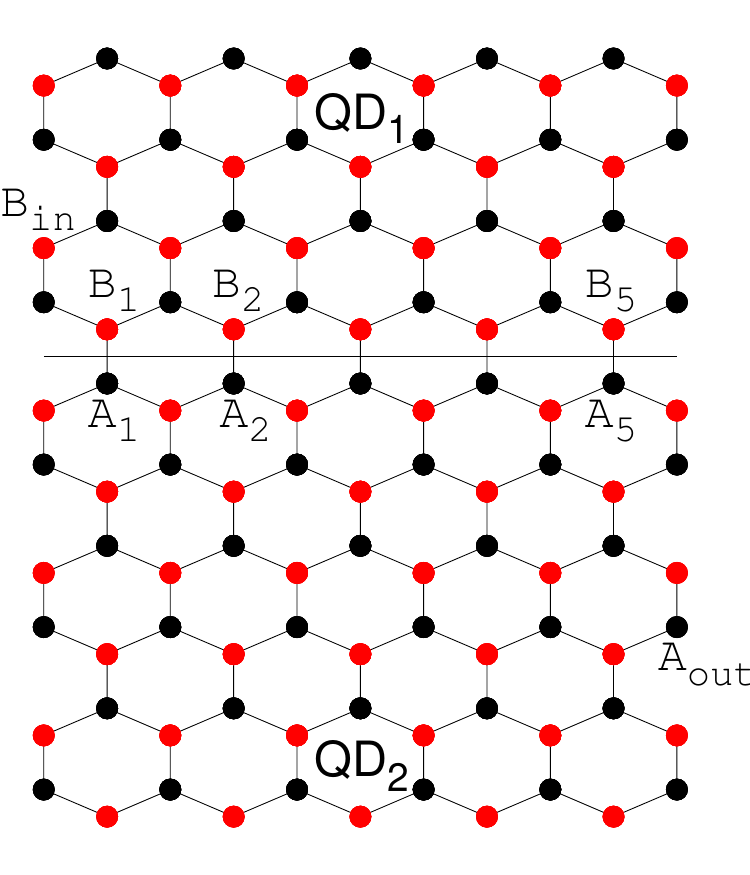}
\caption{
A picture of a graphene explaining the existence of zero transmission
$t_{A_{out}, B_{in}}(0)=0$ for all $B_{in}\in QD_1$ and $A_{out}\in QD_2$ for any constrinction $\tau_c$.
}
\label{grafena2}
\end{figure}

%
%
%
In this section we study the existence
of the ${\bf T}_{AB}$ zeroes for a two-terminal graphene $QD$ at $E=0$.
In Fig.\,\ref{grafena2} the graphene sheet has
the incoming lead connected at the site $i_1=B_{in}$ which belongs to the $B$ sublattices on left arm-chair boundary
and the contact point of the outgoing lead $i_2=A_{out}$ belongs to the $A$ sublattices on the right arm-chair boundary.
In order to apply the above discussed formalism, the graphene is formally separated in two smaller dots $QD_1$ and $QD_2$ that are serially
connected through $N_{zz}$ lines $B_1A_1$,..., $B_5A_5$ that play the role of connection leads between them.
Each smaller dots behaves like a zero conductances device described in Fig.\,\ref{placheta2}.
$QD_1$ has leads connected to the $B$ sublattice and $QD_2$ to the $A$ sublattice.
Both of them have no zero energy eigenstate \cite{malysheva}.
In this instance, Eq.\,\ref{nofanot} applies and the conductance cancels at $E=0$.

From Ref.\onlinecite{malysheva} the rectangular graphene lattice
has pairs of zig-zag edge states $\Psi_{zz+}$ and $\Psi_{zz-}$
with the  wave numbers $\xi_{j}=\pi j / (N_{zz}+1)$ and $\delta_j$
that satisfy the characteristic equation
\begin{eqnarray}\label{eqcaracter}
\sinh \delta_j N_{ac} = 2 \cos {(\xi_j /2)} \sinh \delta_j (N_{ac} + 1/2).
\end{eqnarray}
$N_{zz}$ counts the  zig-zag points and $N_{ac}$ is the number of hexagonal cells in the arm-chair direction.
The two zig-zag states energies are
\begin{eqnarray}\label{ezz}
E_{zz\pm}=\pm \frac  {\sinh(\delta_j /2) } {\sinh \delta_j (N_{ac} + 1/2)}.
\end{eqnarray}
For graphene in Fig.\,\ref{grafena2} we have $N_{zz}=5$ and $N_{ac}=5$.
From Eq.\,\ref{eqcaracter} we obtain only one pair of zig-zag edge states having
the wave numbers $\xi_{5}=5\pi / 6$ and $\delta_5=1.317$.
Their zig-zag energies calculated with Eq.\,\ref{ezz} are $E_{zz\pm}=\pm0.001 t$.
 $t$ is the nearest neighbour hopping {\color{black} equal to 2.7\,eV for nanographene \cite{valli2019}.}

\begin{figure}
\centering
\includegraphics[scale=0.5]{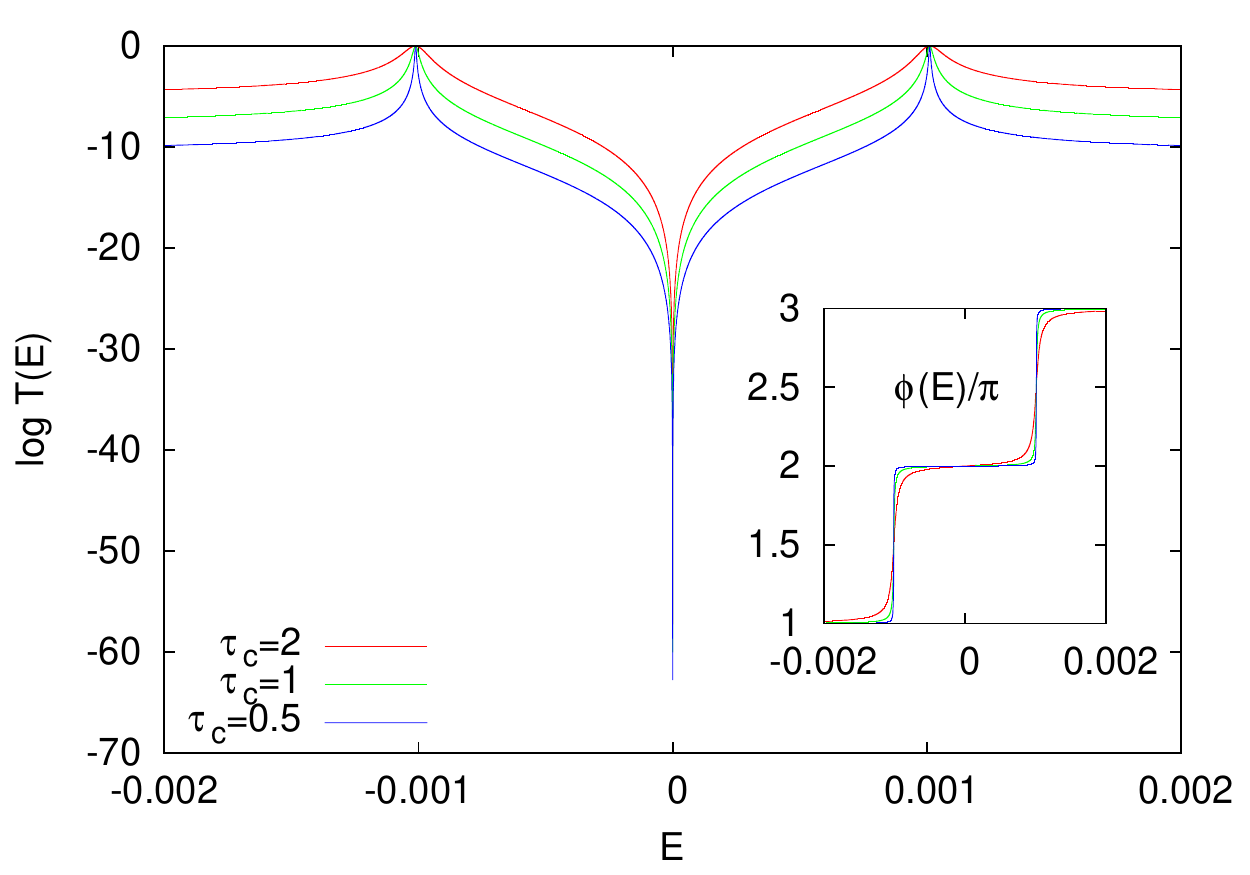}
\caption{
Zero transmission and no phase lapse in Graphene Quantum Dot at $E=0$.
{\color{black} T(0)=0 for any $\tau_c$ is proven in the text.}
The lattice picture and the contact points are in Fig.\,\ref{grafena2}.
$\tau_l=2$.
The lead-dot constrinction parameters $\tau_c$ are written on the figure.
{\color{black} E, $\tau_l$ and $\tau_c$ are in units of t.
In the inset the transmission phase is in $\pi$ units.}
}
\label{figac0}
\end{figure}

In Fig.\,\ref{figac0} we show numerical results of transmittance ${\bf T}(E)$ and the transmission phase $\phi(E)$
when two transport leads are contacted to the points
$A_{out}$ and $B_{in}$ as explained
in Fig.\ref{grafena2}.

{
The maxima with $T(E)=1$ for tunneling energies is obtained at the two resonance energies equal to zig-zag eigenstates calculated above,
$E \simeq E_{zz+}$ and $E\simeq E_{zz-}$.
Between the two resonances the system shows a zero transmittance at $E=0$
with no phase lapse of the transmission phase between them.}
At resonances the phase $\phi(E)$ increases with $\pi$ as expected.

The ${\bf T}_{AB}$ zeros have an increased robustness.
Two different impurities like $\epsilon_B$ located in $QD_1$ and $\epsilon_A$ located in $QD_2$
do not modify the ${\bf T}_{AB}(0)=0$ of Fig.\ref{grafena2}.
In order to lift the conductance zero, one needs at least one
$\epsilon_A$ impurity in $QD_1$ and one $\epsilon_B$ impurity in $QD_2$.
This results from Eq.\,\ref{tpeff} and can be applied to design an AND logical gate by using the graphene QD.
{ In this case the two control parameters $\epsilon_A$ and $\epsilon_B$ can be simmulated
by external perturbations applied on the two selected sites as in the case of B\"uttiker probes \cite{chen2018}.
}

Further, the transmission cancellation proven in this paper can be used to also explain the DQI
in molecular systems that contain a series of subsystems.
If, for instance, the building block is a meta-benzene, we can obtain the zero conductance in biphenyl \cite{solomon2015},
and if we use a T-shape as a building block we obtain the 2-3 hard zero in butadiene \cite{tsuji2017}.
{\color{black}
One may start also with multi-terminal lattices as pictured in Fig.\ref{placheta2}.
As example, a three-terminal naphtalene, with all $T_{AA}=0$,
 may be used as a builiding block to explain the DQI in perylene type lattices
as those obtained in \cite{mayou2013, stuyver2015}.
Finally one remarks that the second dot in Fig.\,\ref{cis} can be choosen arbitrary and in this way one can explain and predict DQI
in more complex systems.
}

\section{Conclusions}

 A large class of molecules and lattices are bipartite, for which this paper addresses a particular transmission cancellation property, potential of use for nano-electronics.

We demonstrate the existence of {\it zero} transmission at half-filling
in bipartite systems, such as graphene quantum dots, when the two transport leads
are contacted to certain sites from the $A$ and $B$ different sublattices. This {\it perfect} transmission cancellation, independent on the coupling strength to the leads, is different from the usual low conductance between resonances, and the property can be used for on/off nano-switches or logical gates. {The algorithm described in this paper is appropriate for bipartite systems that can be separated in two sub-systems, each of them bipartite and lacking mid-spectrum (zero) energy. Then if the two leads are connected to any {\it A} site of the first sub-system and, respectively, to any {\it B} site of the second sub-system, the transmission exhibits a cancellation.}

A high robustness is proven for the ${\bf T}_{AB}$ conductance zeros which survive to any single-site perturbation and at least two impurities (located in different sub-lattices) are necessary to remove them.  This is unlike to the $T_{AA}$ zeros {\color{black} which are invariant to any $A$ site perturbations
and can be lifted by a single $B$ site impurity}.
In addition to the conductance cancellation, no $\pi$ lapse
of the transmission phase occurs if the leads are connected to different sub-lattices,
contrary to the case when the leads are connected to the same sub-lattice. 


Our results can be used to predict the existence of DQIs and to understand their robustness
 in various physical systems
 -as finite tight-binding lattices or molecules- that are composed from various building blocks
with certain bipartite characteristics.

\begin{acknowledgements}
The authors thank Paul Gartner for the help regarding the transport formalism
and Bogdan Ostahie for the useful numerical calculations.
The work was supported by Romanian Core Program PN19-03 (contract no. 21 N/08.02.2019).\end{acknowledgements}

\end{document}